\documentclass[twocolumn,aps,prl,reprint,showpacs,floatfix]{revtex4-2}

\usepackage{color}
\usepackage[utf8]{inputenc}
\usepackage{amsmath,amssymb,amstext}
\usepackage{amsthm}
\usepackage{dsfont}
\usepackage{graphicx}
\usepackage{slashed}
\usepackage{braket}
\usepackage[bookmarksnumbered,pdfpagelabels=true,plainpages=false,colorlinks=true,linkcolor=blue,citecolor=red,urlcolor=blue]{hyperref}
\usepackage[normalem]{ulem}

\newcommand{\del}{\partial}
\renewcommand*{\vec}[1]{\boldsymbol{#1}}

\DeclareMathOperator{\Tr}{Tr}

\newcommand{\SC}{\mathrm{SC}}
\newcommand{\sM}{\mathrm{QH}}
\newcommand{\sigbar}{\bar{\sigma}}
\DeclareMathOperator{\hc}{h.c.}

\begin{document}

\title{Fractionally Charged Vortices at Superconductor-Chern Insulator Interfaces}

\author{Enderalp Yakaboylu}
\email{enderalp.yakaboylu@uni.lu}

\author{Thomas L. Schmidt}
\email{thomas.schmidt@uni.lu}
\affiliation{Department of Physics and Materials Science, University of Luxembourg, L-1511 Luxembourg}

\date{\today}

\begin{abstract}
We investigate the interfacial vortex physics of a heterostructure  composed of a type-II $s$-wave superconductor (SC) and a $C=1$ Chern insulator (CI). By deriving an effective $(2+1)$-dimensional theory, we show that the interfacial Cooper-pair degrees of freedom are described by two coupled Abelian-Higgs fields interacting via a Chern-Simons term inherited from the CI. This interaction endows the photon field with a topological mass and induces a \emph{fractional} electric charge of $e/2$ on the vortices. The topological mass fundamentally reshapes the interfacial vortex lattice, while the fractional charge leads to the formation of unique four-vortex bound clusters. We thus predict a \emph{topological Abrikosov lattice}, establishing a novel phase of matter at the SC-CI interface.
\end{abstract}

\maketitle

\textit{Introduction.} Superconductivity arises from the spontaneous breaking of the global $U(1)$ phase symmetry associated with charge conservation, with the electromagnetic response governed by the Anderson-Higgs mechanism. Starting from the microscopic Bardeen-Cooper-Schrieffer (BCS) theory and integrating out the fermionic degrees of freedom, one obtains the Ginzburg-Landau (GL) theory as an effective field theory for superconductors. In this description the superconducting order parameter is pinned near the minimum of a Higgs-like potential and the photon acquires a mass.

Superconductors (SCs) are classified into two types based on the relationship between two characteristic length scales. The coherence length, $\xi$, is related to the inverse Higgs mass and sets the length scale over which the order parameter varies in space, while the penetration depth, $\lambda$, is related to the inverse photon mass in the broken-symmetry phase and governs the decay of magnetic fields inside the SC. Type-I SCs ($\sqrt{2}\lambda < \xi$) expel magnetic fields and thus exhibit a strong Meissner effect. In contrast, type-II SCs ($\sqrt{2}\lambda > \xi$) allow magnetic flux to penetrate in the form of quantized vortices.

These vortices are well understood in GL theory~\cite{abrikosov1957magnetic,sandier2008vortices,de2018superconductivity} and in its relativistic counterpart, the Abelian Higgs model, where they are electrically neutral~\cite{nielsen1973vortex}. In $(2+1)$ dimensions, however, adding a Chern-Simons (CS) term qualitatively changes their properties by endowing vortices with electric charge~\cite{paul1986charged}. In SC contexts, where the matter field represents Cooper pairs, this induced charge can even become fractional~\cite{hong1990multivortex}.

In this Letter, we propose a condensed-matter realization of charged vortices at the interface between a type-II $s$-wave SC and a $C=1$ Chern insulator (CI), i.e., quantum anomalous Hall phase. While recent experimental progress has successfully coupled superconductors to both quantum Hall states \cite{amet2016supercurrent,lee2017inducing,zhao2020interference,gul2022andreev,hatefipour2022induced,vignaud2023evidence} and quantum anomalous Hall systems \cite{uday2024}, the specific vortex dynamics at these interfaces remain largely unexplored. We demonstrate that the low-energy interface theory is governed by two coupled Abelian-Higgs fields representing interfacial Cooper-pair dynamics, along with an emergent CS term originated from the CI sector. 

The CS term generates a topological photon mass that renormalizes the effective penetration depth, allowing the interface to be tuned between type-II and type-I regimes. Moreover, the CS term endows vortices with a fractional electric charge of $e/2$, favoring four-vortex bound clusters whose total charge equals that of a Cooper pair. This distinguishes the interfacial vortices from both conventional superconducting vortices and those in the standard CS Abelian-Higgs model.

We analyze the SC-CI interface within a framework of quantum field theory. The CI is modeled in $(2+1)$-dimensional spacetime with metric $g^{\mu\nu}=\mathrm{diag}(+,-,-)$, while the SC lives in $(3+1)$ dimensions (D). We use natural units ($\hbar=c=1$) and the Feynman slash notation $\slashed{A}=\gamma^\mu A_\mu$, with the $\gamma$-matrices defined in Supplemental Material (SM). For a spinor $\psi$ we denote the transpose by $\psi^T$ and the Dirac adjoint by $\bar{\psi}=\psi^\dagger\gamma^0$; for a matrix $\xi$ we define $\bar{\xi}=\gamma^0\xi^\dagger\gamma^0$.

\textit{Effective Action.} We consider the following total action for the combined SC-CI system,
\begin{equation}
\label{gen_action}
S = S_{\rm CI}+S_{\rm SC}+S_{\rm int}+S_{\rm Maxwell}\, .
\end{equation}
Our effective field theory is independent of microscopic details, but for concreteness we consider a 2D lattice system realizing a $C=1$ CI, which is the lattice realization of the quantum anomalous Hall effect. Its low-energy Hamiltonian is given by $H_{\rm CI}= \left( v_F \vec{k}\cdot\vec{\sigma}+m_q v_F^2\sigma_z \right)$, where $v_F$ is the Fermi velocity,  $\vec{k} = (k_x,k_y)$ is the momentum, and $(\sigma_x,\sigma_y,\sigma_z)$ are the Pauli matrices. The mass term $m_q v_F^2\sigma_z$ breaks time-reversal symmetry and opens a gap, $\epsilon_q(\vec{k})=\pm\sqrt{(v_F\vec{k})^2+(m_q v_F^2)^2}$. Applying minimal coupling to the electromagnetic field and taking the continuum limit, we obtain the $(2+1)$D action for a massive Dirac field $\psi (t,x,y)$ coupled to the classical vector potential $A_\mu(t,x,y)$,
\begin{equation}
S_{\rm CI}=\int d^3x \, \bar{\psi}_q \left(i\slashed{D}-m_q v_F^2\right)\psi_q\, ,
\end{equation}
where we parametrize the field as $\psi_q = v_F\,\psi (t,v_F x,v_F y)$. This (unconventional) choice enables a covariant notation across the electronic and photonic sectors despite $v_F\ll c$, using $D_\mu=\partial_\mu+i e A_\mu$ for charge $e$.

To describe the SC sector microscopically, one can in principle start from the standard BCS action~\cite{aitchison1995effective}. However, since the CI sector is written in relativistic form, it is convenient to approximate the BCS action by a relativistic effective theory. We therefore introduce the following $(3+1)$D action for the Dirac field $\psi_s = v_F^{3/2}\psi (t,v_F x,v_F y,v_F z)$ for the SC,
\begin{align}
\label{sc_action}
S_{\rm SC}=\int d^4x\,\bigg[\, &\bar{\psi}_s\left(i\slashed{D}-\tilde{m}_s v_F^2+m_s v_F^2\gamma^0\right)\psi_s \notag\\
&+g\,(\psi_s^T\xi_s\psi_s)\,(\bar{\psi}_s\bar{\xi}_s\bar{\psi}_s^T)\bigg]\, .
\end{align}
Here, $\tilde{m}_s v_F^2 = m_s v_F^2\sqrt{1-2\mu_s/(v_F^2 m_s)}$, with $\mu_s$ the chemical potential. For simplicity we take the Fermi velocities to be identical in the SC and CI sectors. The last term represents an attractive interaction of strength $g > 0$ that drives Cooper pairing below a critical temperature. Choosing $\xi_s = i\sigma_y/2 \otimes I_2$ yields $s$-wave pairing at the mean-field level~\cite{capelle1999relativistic,ohsaku2001bcs}. The corresponding quasiparticle dispersion can be written as $\epsilon_s(\vec{k})=\sqrt{(v_F\vec{k})^2-(v_F k_F)^2+(m_s v_F^2)^2}-m_s v_F^2$, which reduces to $|\epsilon_s(\vec{k})|\approx \vec{k}^2/(2m_s)-\mu_s$ with $k_F=\sqrt{2m_s\mu_s}$ in the limit $\sqrt{|k^2-k_F^2|}\ll m_s v_F$. Hence, in this regime Eq.~\eqref{sc_action} reproduces the standard non-relativistic BCS action.

The third term in Eq.~\eqref{gen_action} describes the interaction between SC and CI electrons. Coupling to an $s$-wave SC allows Cooper pairs to tunnel into the CI surface states via the proximity effect~\cite{fisher1994cooper}. This is captured by the pair-pair interaction
\begin{equation}
\label{int_action}
S_{\rm int}=\eta\int d^4x\,\delta(z)\left[(\psi_s^T\xi_s\psi_s)(\bar{\psi}_q\bar{\xi}_q\bar{\psi}_q^T)+\text{h.c.}\right],
\end{equation}
with effective strength $\eta > 0$, typically small compared to the bare pairing, $\eta \ll g$. Such a term can be derived via a Schrieffer-Wolff transformation starting from single-electron tunneling; parametrically, $\eta \sim g\,\gamma_T^2/\Delta^2$, where $\gamma_T$ is the tunneling amplitude and $\Delta$ the superconducting gap (see SM). Here $\xi_q = i\sigma_y/2$ is the pairing matrix in the CI system, and the Dirac delta confines the interaction to the $x$-$y$ interface plane at $z=0$. Finally, the last term in Eq.~\eqref{gen_action} is the Maxwell action,
\begin{equation}
S_{\rm Maxwell}=-\frac{1}{4}\int d^4x\,F_{\mu\nu}F^{\mu\nu},
\end{equation}
with field-strength tensor $F_{\mu\nu}=\partial_\mu A_\nu-\partial_\nu A_\mu$.

To derive an effective theory for the interface, we consider the path integral
$Z[A]=\int D(\bar{\psi}_s,\psi_s)\,D(\bar{\psi}_q,\psi_q)\,e^{iS}$,
assuming translational invariance along the $z$ direction in the SC sector~\cite{sandier2008vortices}. We first decouple the quartic pairing interactions in Eqs.~\eqref{sc_action} and \eqref{int_action} by introducing complex Hubbard-Stratonovich fields $\phi_1$ and $\phi_2$, which at the mean-field level describe Cooper-pairs with charge $q = 2e$ in the SC and CI sectors, respectively, and then integrate out the fermions (see SM for details). This yields the bosonic functional
\begin{equation}
Z[A]=\int D(\bar{\phi}_1,\phi_1)\,D(\bar{\phi}_2,\phi_2)\,e^{iS_{\rm HCS}}\,,
\end{equation}
with the $(2+1)$D Higgs-Chern-Simons (HCS) action
\begin{align}
\label{eff_action}
& S_{\rm HCS}
= \frac{k}{4\pi}\int d^3x\,\varepsilon^{\mu\nu\rho}A_\mu\partial_\nu A_\rho
-\frac{1}{4}\int d^3x\,F_{\mu\nu}F^{\mu\nu}\notag \\
&+\int d^3x\,\sum_{j=1}^2\left[\left|(\partial_\mu+i q A_\mu)\phi_j\right|^2
-\nu_j\left(|\phi_j|^2-\delta_j^2/2\right)^2\right]\notag\\
& - \beta\int d^3x\,\left(\bar{\phi}_2\,\phi_1+\text{h.c.}\right)\, .
\end{align}
The first term in this action is a CS term generated by the one-loop polarization of the CI fermions. The coefficient $k$ sets the CS level parameter, $k/e^2=1$, corresponding to $C=1$. This integer value is obtained by integrating out the fermions together with a heavy Pauli-Villars (PV) regulator field, which provides a proper UV completion of the Dirac theory and respects the fermion-doubling structure of an underlying lattice model. Without such a UV regularization one would instead obtain an unphysical half-integer CS level.

The second line represents a GL (or Abelian-Higgs, in high-energy terminology) action for each sector. The parameters $\nu_j>0$ and $\delta_j>0$ arise from a perturbative expansion of the interaction terms and can be expressed in terms of fermionic loop integrals. In particular, $\delta_1$ characterizes the superconducting gap in the SC sector, while $\delta_2$ is generated by the pair-pair coupling and represents the induced gap in the CI layer. In the absence of a chemical potential, the coefficient of the quartic term $|\phi_2|^4$ vanishes, rendering $\delta_2$ ill-defined, since the density of states in the CI sector is zero inside the gap. Introducing a nonzero chemical potential $\mu_q \ge m_q v_F^2$ shifts the Fermi level, and  thereby generates a nonzero $|\phi_2|^4$ term. In particular, for $\mu_q = m_q v_F^2$ one obtains
\begin{equation}
\delta_2=\sqrt{8\pi m_q v_F^2\left(g L_{\rm CI}/\eta^2- m_q v_F^2/(2\pi) \right)}\, ,
\end{equation}
where $L_{\rm CI}$ denotes the thickness of the CI layer. Notet that for this choice of $\mu_q$ the longitudinal Drude current, expressed as $(\mu_q^2-(m_q v_F^2)^2)/\mu_q$, vanishes so the system remains in the insulating regime.

The last term in Eq.~\eqref{eff_action} represents an interfacial Josephson coupling between the two pair fields, which energetically favors a fixed relative phase (phase locking) across the interface. Within our perturbative treatment, its coupling strength is given by $\beta = L_{\rm CI}/\eta$.

\textit{Spontaneous symmetry breaking.} In the absence of the CI sector, i.e., for $\phi_2 = k = \beta = 0$, the field $\phi_1$ condenses by settling into the minimum of the Higgs potential $\bigl(|\phi_1|^2-\delta_1^2/2\bigr)^2$, thereby spontaneously breaking the $U(1)$ symmetry and forming a superconducting state. At the mean-field level this distinguishes the superconducting phase, $|\phi_1| = \delta_1/\sqrt{2}$, from the normal phase, $\phi_1 = 0$. In the former phase one defines the Landau parameter $\kappa = \lambda/\xi$, with the magnetic penetration depth $\lambda = \sqrt{2}/(q\delta_1)$ and the coherence length $\xi = 1/(\delta_1\sqrt{\nu_1})$. Values $\kappa < 1/\sqrt{2}$ correspond to type-I SCs, while $\kappa > 1/\sqrt{2}$ imply type-II SCs. In the latter case, the pair $(\phi_1,A^\mu)$ admits vortex solutions, which minimize the energy functional in the limit $r \to\infty$, and magnetic flux penetrates in the form of vortex lines~\cite{coleman2015introduction,dunne2002aspects}.

The full HCS action~\eqref{eff_action} can be analyzed similarly. We expand the fields $\phi_j$ about their vacuum expectation values ($j=1,2$) as
\begin{equation}
\label{vac_expansion}
\phi_j = \frac{1}{\sqrt{2}} (\delta_j + H_j(x)) e^{i \Theta_j (x)} \, ,
\end{equation}
and introduce the linear combinations of the phases
\begin{equation}
    G = \frac{\delta_1^2 \Theta_1 + \delta_2^2 \Theta_2}{\delta^2}, \qquad P = \frac{\delta_1 \, \delta_2}{\delta} (\Theta_1 - \Theta_2) \, ,
\end{equation}
with $\delta^2 = \delta_1^2 + \delta_2^2$. Here, $H_{1,2}(x)$ are the amplitude (Higgs) modes, $G(x)$ is the overall-phase (Goldstone) mode, and $P(x)$ is the relative-phase (Leggett) mode. By performing a gauge transformation $A_\mu \to A_\mu - \partial_\mu G / q$, known as the unitary gauge, we remove the massless Goldstone field from the spectrum. The HCS Lagrangian density then takes the form (see SM)
\begin{align}
& \mathcal{L}_{\rm HCS} = \sum_{j=1}^2 \left[ \frac{(\del_\mu H_j)^2}{2}  - \frac{m_{H_j}^2 H_j^2}{2}  \right] + \frac{(\del_\mu P)^2}{2} - \frac{m_P^2 P^2 }{2} \notag \\
 &  -  \frac{1}{4} F_{\mu \nu}F^{\mu \nu} +  \frac{k}{4 \pi} \varepsilon^{\mu \nu \rho} A_\mu \del_\nu A_\rho
+\frac{1}{2} q^2 \delta^2 A_\mu A^\mu  + \mathcal{L}_{\rm HCS}^{\rm int} \notag  \, ,
\end{align}
where $\mathcal{L}_{\rm HCS}^{\rm int}$ collects interaction terms among the various modes. The corresponding masses are $m_{H_j} = \delta_j \sqrt{2 \nu_j} $, $m_{P} = \delta \sqrt{\beta/(\delta_1 \delta_2)}$, and, for the gauge sector, diagonalizing the quadratic terms yields two massive modes
\begin{equation}
m_{\pm} = \sqrt{q^2 \delta^2 + k^2/16 \pi^2} \pm k/4 \pi \, .
\end{equation}
We note that in the absence of the Higgs sectors ($\delta \to 0$), one finds $m_+ = k / (2\pi)$, known as the topological mass. Thus, the CS term shifts the photon mass from the purely Higgs value $q\delta$ to the hybrid values $m_{\pm}$~\cite{paul1986self}.

These masses set the characteristic length scales extracted from the corresponding equations of motion. In the rotationally symmetric case, the radial profiles decay as $H_j \sim \exp(-m_{H_j}r / \sqrt{2})$ and $A^\nu \sim \exp(-m_\pm r)$. Hence the coherence lengths read
$\tilde{\xi}_j=\sqrt{2}/m_{H_j}=(\delta_j\sqrt{\nu_j})^{-1}$,
while the magnetic penetration depth is given by
\begin{equation}
\tilde{\lambda}=\frac{1}{m_{-}}
=\frac{1}{\sqrt{q^2 \delta^2 + k^2/16\pi^2}-k/4\pi}\, .
\end{equation}
We choose $m_{-}$ rather than $m_{+}$ because the longer penetration depth corresponds to the energetically favorable vortex solution considered later~\cite{paul1986charged}.

The Leggett mode, on the other hand, introduces an additional length scale: the Leggett healing length $\xi_{\rm L}=1/m_P$. This mode, which is a unique feature of the coupled system, can be interpreted as internal degree of freedom in which one condensate oscillates relative to the other, and $\xi_{\rm L}$ controls relaxation of the \emph{relative} phase $\Theta_1-\Theta_2$. In what follows, to obtain admissible vortex solutions, we work in the weak-locking regime, where $\xi_{\rm L}$ is parametrically larger than the other characteristic lengths, i.e., $\xi_{\rm L} \gg \tilde{\lambda}, \tilde{\xi}_j$. Such a regime is realized for a sufficiently thin CI layer as $\beta=L_{\rm CI}/\eta$.

We then introduce the analogous Landau parameters,
$\tilde{\kappa}_j=\tilde{\lambda}/\tilde{\xi}_j$,
which characterize the interfacial superconducting response via $\tilde{\kappa}_j \gtrless 1/\sqrt{2}$. A key distinction from conventional GL theory is that the CS term enters $\tilde{\lambda}$: while $\delta_2$, controlled by the interfacial coupling $\eta$, can be suppressed, the contributions proportional to $k$ persist even for small $\eta$. This reflects the topological origin of the CS term and implies that it can strongly renormalize the penetration depth, and hence the type of superconductivity at the interface, as we show below.

\textit{Vortex Solutions.} In order to seek vortex solutions for the SC-CI interface, we adopt the standard cylindrically symmetric ansatz with winding numbers $n_1, n_2 \in\mathbb Z$, in which the electromagnetic fields depend only on the radial coordinate:
\begin{equation}
\label{vortex_ansatz}
\phi_j = e^{i n_j \theta} f_j (r)\, , \ A^\mu = (a_0 (r), 0, A_\theta  = a(r)/r) \, .
\end{equation}
For the energy to remain finite as $r \to\infty$, the scalars fields approach their vacuum expectation values $ \lim_{r \to \infty} f_j (r)  = \delta_j/ \sqrt{2} $, while  $ \lim_{r \to \infty} a_0 (r) = 0 $ and  $ \lim_{r \to \infty} a(r)  = n_1/q = n_2/q$. The latter enforces the compatibility condition $ n_1=n_2\equiv n \in\mathbb Z$. Regularity at the core instead requires $ \lim_{r \to 0} f_j (r)  =0 $ and $ \lim_{r \to 0} a (r) = 0$, whereas $\lim_{r \to 0} a_0 (r) = a_0 $ is a constant. This constant can be chosen so that the electric field remains regular (and, in particular, vanishes) as $ r\to 0$. Moreover, the magnetic flux is $\Phi_n = - 2 \pi n/ q$, and finite-energy solutions are therefore classified by their winding number $n$. Finally, in the absence of the CI sector (implying $a_0 =0 $), the ansatz reduces to that of the Nielsen-Olesen vortex~\cite{nielsen1973vortex}.

In contrast to electrically neutral Nielsen-Olesen vortices, when the CS term is present, the vortex binds electric charge. This follows from the corresponding equation of motion (the modified Gauss law), $j^0 = \vec{\nabla} \cdot \vec{E} + k B /(2\pi)$, which defines the charge:
\begin{equation}
 Q  = \int d^2 x \, j^0 = \frac{k}{2\pi}  \Phi_n  = -n\, k/q =  - n \, e/2 \, .
\end{equation}
This leads to the crucial conclusion that for an odd winding number, the vortex carries a fractional electric charge despite the underlying integer Hall response. This fractional charge emerges from CS flux-charge attachment for Cooper pairs with $q=2e$. 
One may also view the resulting $e/2$ charge as twice the $e/4$ charge per surface arising in the half-quantized Hall conductivity setup governed by axion electrodynamics and the Witten effect~\cite{witten1979dyons,Fu_2008,nogueira2016josephson,de2025manipulating}. It also aligns with recent indications that the integer quantum Hall regime can provide a robust platform for anyons~\cite{glidic2024signature}.
% This result aligns with recent indications that the integer quantum Hall regime can provide a robust platform for anyons~\cite{glidic2024signature}.

\begin{figure}[t]
  \includegraphics[width=0.8\linewidth]{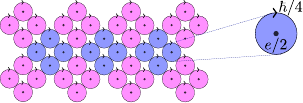}
  \vspace{-0.2cm}
\caption{Illustration of vortex lattice formation in quadruplets.}
\label{fig_vortex}
\vspace{-0.3cm}
\end{figure}

We highlight that the charge can be related to the current via $Q = j_\theta A / v_F $, where $A$ is the interface area, and the azimuthal current density is
\begin{equation}
\label{current}
 j_\theta = - \frac{k}{2 \pi} E_r + (\vec{\nabla} \times \vec{B})_\theta \,.
\end{equation}
The first term corresponds to the CI edge current, while the second term is the London contribution, i.e., the supercurrent on the SC side. Therefore, the vortex charge can in principle be inferred from current measurements. We note that while our focus is on the interfacial bulk, Eq.~\eqref{current} also encodes the corresponding edge behavior.

We further calculate the angular momentum carried by the vortices from the energy-momentum tensor $T_{\mu \nu}$,
\begin{equation}
    M_z = \int d^2 x \, \epsilon^{i j} x_i T_{0 j} =  - \frac{n}{q} Q = \left(\frac{n}{2} \right)^2 \, .
\end{equation}
Since the effective action $S_{\rm HCS}$ describes a bosonic system, its total angular momentum must be an integer. Accordingly, vortices with an odd winding number $n$ should emerge in quadruplets, each consisting of four vortices, so that the combined configuration carries integer angular momentum (see Fig.~\ref{fig_vortex}).

In connection with the formation of vortex lattices in such quadruplets, we also predict an increase in the lattice period at the interface. This effect arises from the enhanced penetration depth, which itself is a consequence of the topological mass term. The increase in lattice spacing can also be understood from vortex interactions: in addition to the usual repulsive interaction energy $\Delta \mathcal{E}(0)$, the vortices experience an additional repulsive contribution due to their electric charge. In order to illustrate the qualitative behavior of the vortex-vortex interaction, we neglect the Josephson coupling and expand the vortex ansatz as a series in $k/(2\pi)$, as shown in the SM. One then finds that the interaction energy between two vortices acquires a topological correction,
\begin{equation}
    \Delta \mathcal{E} (k)  = \Delta \mathcal{E} (0) + \frac{(n^2 - n)}{8 q^2}  \left(\frac{k}{2\pi} \right)^2 + O \left(\frac{k}{2\pi} \right)^4 \, .
\end{equation}
For $n \ge 2$ this correction is positive, enhancing the effective repulsion and thus favoring a larger intervortex spacing at the interface.

\begin{figure}[t]
\centering
\includegraphics[width=0.85\linewidth]{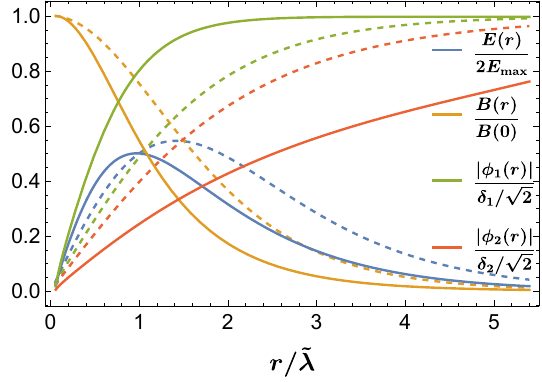}
\vspace{-0.2cm}
\caption{Numerical vortex solutions~\eqref{vortex_ansatz} for $n=1$, $e=1$, and $\beta = 0.05$. Parameters are $\sqrt{2}\tilde{\kappa}_1=2.23$, $\sqrt{2}\tilde{\kappa}_2=0.80$, $\sqrt{2}\kappa=2.23$, $\delta_2/\delta_1=0.4$ for solid lines, and $\sqrt{2}\tilde{\kappa}_1=0.85$, $\sqrt{2}\tilde{\kappa}_2=0.55$, $\sqrt{2}\kappa=1.02$, $\delta_2/\delta_1=0.66$ for dashed lines.}
\label{fig_num}
\vspace{-0.3  cm}
\end{figure}

Finally, the CS-induced renormalization of the penetration depth can change the effective superconducting type at the interface. In Fig.~\ref{fig_num} we show numerical solutions of the vortex equations, obtained by solving the Euler-Lagrange equations of the action $S_{\rm HCS}$ for the ansatz~\eqref{vortex_ansatz} in the regime where $\beta \ll \delta_i$. For the solid curves, the superconducting pair field remains in the type-II regime, with $\sqrt{2}\,\tilde{\kappa}_1 = 2.23$, while the induced CI pair field is type-I, with $\sqrt{2}\,\tilde{\kappa}_2 = 0.80$. By contrast, the dashed curves correspond to a regime in which both fields are type-I at the interface, even though the \emph{bulk} superconductor is type-II. Notably, our numerical calculations indicate that no vortex solutions exist for $\beta \sim \delta_i$.

Regarding experimental tests of our predictions, a concrete realization of the CI is provided by magnetic topological insulator (MTI) thin films based on the Bi-Te or Bi-Se family of materials. In the presence of a finite magnetization, such films enter the quantum anomalous Hall regime and are well described at low energies by a massive Dirac Hamiltonian with Chern number $C=1$~\cite{qi2006,yu2010,wangQuantumAnomalousHall2013,zsurka2024low,legendre2024}. 
In our proposal, the applied magnetic field will cause an orbital effect on the electrons in this quantum anomalous Hall state without closing the gap and drive the SC into the mixed state. Moreover, it has already been demonstrated experimentally that superconductivity can be induced in MTI thin films via the superconducting proximity effect~\cite{uday2024}. This makes the SC-MTI heterostructure a natural platform to test our predictions. In addition, scanning SQUID microscopy is by now a standard technique to detect magnetic-flux vortex lattices and has been used to probe edge currents in MTIs~\cite{ferguson2023}. This suggests that the interfacial vortex lattice and the associated fractional bound charge should, in principle, be accessible to experimental detection.

\textit{Conclusion.} In summary, we have studied the interface between a type-II $s$-wave SC and a $C=1$ CI. We showed that the interfacial physics is captured by two Abelian-Higgs fields, associated with the interfacial Cooper pairs, coupled to a gauge field with both a Maxwell term and an emergent CS term. The latter qualitatively modifies the Abrikosov lattice at the interface: it generates a topological contribution to the photon mass, thereby renormalizing the penetration depth and enabling a change in the effective superconducting type; moreover, through its coupling to Cooper pairs, it endows interfacial vortices with fractional electric charge.

\begin{acknowledgments}
The authors are grateful to Wolfgang Belzig, Michele Governale, and Uli Z\"ulicke for insightful discussions. TLS acknowledges support by the National Research Fund Luxembourg under Grant No.~INTER/QUANTERA21/16447820/MAGMA.
\end{acknowledgments}

\vspace{-0.5cm}

\bibliography{bib_vortex}

\clearpage
\onecolumngrid
\appendix

\setcounter{equation}{0}

\section{Supplemental Material}

\subsection{Emergence of the Chern-Simons Term}

We consider the simplest two-dimensional lattice system realizing a $C=1$ Chern insulator (CI), i.e., the lattice analog of the quantum anomalous Hall effect, whose low-energy Hamiltonian is
\begin{align}
    H_{\rm CI} &= \sum_{\vec{k}} C_{\vec{k}}^\dagger \begin{pmatrix} m_q v_F^2 & v_F(k_x - i k_y) \\ v_F(k_x + i k_y) & - m_q v_F^2 \end{pmatrix} C_{\vec{k}}\, , \qquad  \text{where} \quad C_{\vec{k}} = \begin{pmatrix} c_{\vec{k} \uparrow} \\ c_{\vec{k} \downarrow} \end{pmatrix} \, .
\end{align}

In the continuum limit, this system can be described by the action of a Dirac field $\psi (t,x,y)$ of mass $ m_q v_F^2$
\begin{equation}
\label{dirac_action}
  S_{\rm CI}  = \int dt \, d^2 x \, \bar{\psi} (t,x,y) \left( i \gamma^0 \del_t + i v_F \gamma^i \del_i  - m_q v_F^2 \right) \psi (t,x,y) \, ,
\end{equation}
where $\gamma^0 = \sigma_z$ and $\gamma^{1,2} = i \sigma_{x,y}$ denote the gamma matrices in 2+1 dimensions (D). The energy can be written as $\epsilon_q (\vec{k})= \pm \sqrt{(v_F \vec{k})^2 + (m_q v_F^2)^2}$. In the limit of $m_q \to 0$ one finds a Dirac cone with $\epsilon_q (\vec{k}) = \pm v_F |\vec{k}|$.

The Dirac action~\eqref{dirac_action} can be further expressed as
\begin{align}
S_{\rm CI} &  =  \int c dt \, d^2 x \, \bar{\psi} (t,x,y) \left( i \gamma^0 \del_0 + i \frac{v_F}{c} \gamma^i \del_i  - m_q \frac{v_F^2}{c} \right) \psi (t,x,y) \notag \\
& = \frac{v_F^2}{c^2} \int d^3 x' \, \bar{\psi} (t,x' \, v_F /c, y' \, v_F /c) \left( i \gamma^0 \del_0 + i \gamma^i \del'_i  - m_q \frac{v_F^2}{c} \right) \psi (t, x' \, v_F/c, y' \, v_F / c) \notag \\
& =  \int d^3 x \, \bar{\psi}_q (t,x,y) (i \gamma^\mu \del_\mu - m_q v_F^2) \psi_q (t,x,y)\, .
\end{align}
Here, we define $\psi_q (t,x,y) =  (v_F/c) \, \psi (t, x \, v_F/c, y \, v_F / c) $ and set $c=1$. The Dirac field can be expanded in terms of Grassmann variables $\eta_n$ as
\begin{equation}
\psi_q (t,x,y) = \sum_n \eta_n \chi_n (t,x,y) \, ,
\end{equation}
where $\chi_n (t,x,y)$ denote a complete basis of the eigenstates of the Dirac equation in (2+1)D. Furthermore, after we couple the Dirac field to a $U(1)$ gauge field $A_\mu (t,x,y)$, we redefine the action describing the CI sector as
\begin{equation}
S_{\rm CI} = \int d^3 x \, \bar{\psi}_q \left( i \slashed{\del} - e \slashed{A} - m_q v_F^2 \right) \psi_q \, .
\end{equation}

The corresponding path integral is given by
\begin{equation}
Z[A]=\int D (\bar{\psi}_q,\psi_q) e^{i S_{\rm CI}} = \int D (\bar{\psi}_q,\psi_q) e^{i \int d^3 x\, \bar{\psi}_q M \psi_q} = N \det M \, ,
\end{equation}
where $N$ is a normalization factor (see below), and 
\begin{equation}
M =  i \slashed{\del} - e \slashed{A} - m_q v_F^2 = G_0^{-1} (1 + i e G_0 \slashed{A}) \, .
\end{equation}
Here, $G_0^{-1} =  i \slashed{\del}  - m_q v_F^2$ denotes the differential operator of the unperturbed massive Dirac field, and $G_0$ is the Feynman propagator, i.e., the time-ordered Green's function, given by
\begin{equation}
G_0  (x-y) = \int d^3 p \, G_F (p) e^{-i p (x-y)}, \qquad G_F(p) =  \frac{i(\slashed{p}+m_q v_F^2)}{p^2 - (m_q v_F^2)^2 +i 0^+} \, .
\end{equation}
We use the convention that $ - i G_0^{-1}\, G_0 = 1$, which follows from
\begin{equation}
\left( i \slashed{\del}_x - m_q v_F^2 \right)_{\alpha \beta}  [G_0 (x-y)]_{\beta \gamma} = i \delta_{\alpha \gamma} \delta^{(3)} (x-y) \, .
\end{equation}
By choosing the normalization as $N = (\det G_0^{-1})^{-1}$ such that $Z[0]=1$, the path integral can be written as
\begin{equation}
Z[A] = \exp\left( - \sum_{n=1}^\infty \frac{1}{n} \Tr (-i e G_0 \slashed{A} )^n \right) \, ,
\end{equation}
where the trace can be evaluated via 
\begin{equation}
\Tr (- i e G_0 \slashed{A} )^n = (-i e)^n \int d^3 x_1 d^3 x_2 \cdots d^3 x_n  \Tr \slashed{A} (x_1) G_0 (x_2 -x_1) \slashed{A}(x_2) G_0 (x_3 - x_2) \cdots \slashed{A}(x_n) G_0 (x_1 - x_n) \,.
\end{equation}
By defining $Z[A] = \exp(i S_{\rm eff}[A])$, the leading term of the effective action $S_{\rm eff}[A]$ can be expressed as
\begin{align}
i S_{\rm eff}[A] & = \frac{e^2}{2} \int d^3 x_1 d^3 x_2 \Tr \slashed{A}(x_1) G_0 (x_2 -x_1) \slashed{A}(x_2) G_0 (x_1 - x_2) 
\label{eff_action_general_fer}
 =  \frac{e^2}{2} \int d^3 p \,  \tilde{A}_\mu (-p)  \tilde{A}_\nu (p) \Pi^{\mu \nu} (p) \, ,
\end{align}
where $\tilde{A}_\mu $ is the Fourier transform of the gauge field $A_\mu$ and $\Pi^{\mu \nu}$ is the polarization tensor arising from the one-loop Feynman diagram
\begin{align}
\label{polarization_tensor}
    \Pi^{\mu \nu} (p)
&= 
   \int d^3 k \,  \Tr \gamma^\mu G_F (k-p )\gamma^\nu G_F(k)  \notag \\
&= 
    -2i \varepsilon^{\mu\rho \nu} p_\rho \int d^3 k \frac{m_q v_F^2}{((k-p)^2-(m_q v_F^2)^2+ i 0^+)(k^2-(m_q v_F^2)^2 + i 0^+)} + \text{divergent terms}\, ,
\end{align}
where we use $\Tr(\gamma^\mu \gamma^\nu \gamma^\rho ) = - 2i \epsilon^{\mu \nu \rho}$. In the limit of $m_q v_F^2 \gg |p|$ the $k$-integral can be written as
\begin{equation}
\label{polarization_no_chemical}
\int d^3 k  \frac{2 m_q v_F^2}{(k^2- (m_q v_F^2)^2 + i 0^+)^2} = i \int d^3 k_E  \frac{2 m_q v_F^2}{(k_E^2+(m_q v_F^2)^2 )^2} = \frac{i}{4\pi} \frac{m_q}{|m_q|} \, ,
\end{equation}
where we performed a Wick rotation $k \to k_E$ and used the identity
\begin{equation}
\int d^D k_E \frac{1}{(k_E^2 + m^2)^n} = \frac{1}{(4\pi)^{D/2}} \frac{\Gamma(n-D/2)}{\Gamma[n]}\frac{1}{(m^2)^{n-D/2}} \, . 
\end{equation}
Therefore, in the limit of $m_q v_F^2 \gg |p|$, the effective action is given by
\begin{equation}
S_{\rm eff} [A, m_q] = \frac{\text{sign}(m_q)}{2} \frac{e^2}{4\pi}\int d^3 x  \, \varepsilon^{\mu  \rho \nu} A_\mu \del_\rho A_\nu + \text{divergent terms} \, .
\end{equation}
The divergent terms, on the other hand, can be dealt with by introducing a suitable regularization scheme in the UV sector. Particularly, in the Pauli-Villars (PV) scheme, the regularized effective action can be defined as
\begin{equation}
S_{\rm eff}^{\rm PV\pm}[A] = S_{\rm eff} [A, m_q] - \lim_{\Lambda \to \pm \infty} S_{\rm eff} [A, \Lambda] \, .
\end{equation}
Within the framework of the negative-mass PV regularization scheme, the regularized action is given by
\begin{equation}
S_{\rm CS} = S_{\rm eff}^{\rm PV-} =  \frac{k}{4\pi}\int d^3 x \,  \varepsilon^{\mu \rho \nu} A_\mu \del_\rho A_\nu \, ,
\end{equation}
which is the Chern-Simons (CS) action with the level parameter $k/e^2 = 1$. This shows that the CI sector contributes a CS term to the effective action.

\subsection{Emergence of the Ginzburg-Landau Term}

As our main goal is to explore the interface between the superconductor (SC) and the CI system, and the action of the latter is in the relativistic form, we will approximate the BCS action with its relativistic form for the sake of computational simplicity. First, note that for the kinetic energy of the electrons with mass $m_s$ in the SC, we have $\epsilon_{\rm BCS} (k_F) = 0$ and $\epsilon_{\rm BCS}'(k_F) = k_F/m_s$, where $k_F = \sqrt{2 m_s \mu_s}$ in terms of the chemical potential $\mu_s$. We define the relativistic counterpart of this spectrum as
\begin{align}
    \epsilon_s (\vec{k})= \sqrt{ (v_F \vec{k})^2 - (v_F k_F)^2 + (m_s v_F^2)^2} - m_s v^2_F \, ,
\end{align}
which satisfies the same conditions,
\begin{equation}
 \epsilon_s (k_F) = 0  \, , \quad 
    \epsilon_s'(k_F) = \frac{2 v_F^2 k_F}{2 m_s v_F^2} = \frac{k_F}{m_s} \, .
\end{equation}
Furthermore, in the limit $\sqrt{|\vec{k}^2 - k_F^2|} \ll m_s v_F$, the spectrum reduces to
\begin{align}
  \epsilon_s (\vec{k}) \to \epsilon_{\rm BCS} (\vec{k}) = \frac{\vec{k}^2}{2m_s} - \mu_s \, .
\end{align}
Therefore, we first consider the following Dirac action
\begin{equation}
\label{dirac_action_SC}
  S_{\rm Dirac} = \int dt \, d^3 x \, \bar{\psi} (t,x,y,z) \left( i \gamma^0 \del_t + i v_F \gamma^i \del_i  - \tilde{m}_s v_F^2 +m_s v_F^2 \gamma^0 \right) \psi (t,x,y,z) \, ,
\end{equation}
where $\tilde{m}_s v_F^2 = m_s v_F^2 \sqrt{1 - 2 \mu/(m_s v_F^2 ) }$ and the gamma matrices in a $3+1$ dimensional spacetime are given by
\begin{equation}
\gamma^0 = \begin{pmatrix}
   I_2 & 0 \\
   0 & - I_2
\end{pmatrix} \, , \quad \gamma^i =\begin{pmatrix}
  0 &  \sigma^i \\
   - \sigma^i & 0
\end{pmatrix}  \, ,
\end{equation}
with $I_2$ being the 2-dimensional identity matrix. Similarly to the CI sector, we can rewrite the Dirac action~\eqref{dirac_action_SC} as
\begin{equation}
    S_{\rm Dirac} = \int d^4 x \, \bar{\psi}_s \left( i \slashed{\del} -  \tilde{m}_s v_F^2 +m_s v_F^2 \gamma^0 \right) \psi_s \, ,
\end{equation}
where $\psi_s (t,x,y,z) = (v_F)^{3/2} \psi (t,x \, v_F,y\, v_F,z\, v_F) $, and it can be expanded as
\begin{equation}
\psi_s (t,x,y,z) = \sum_n \eta_n \chi_n (t,x,y,z) \, ,
\end{equation}
with $\chi_n (t,x,y,z)$ being a basis of 4-component spinors.

By further taking into account the four-fermion interaction, we introduce the following relativistic BCS action
\begin{equation}
\label{BCS_action}
S_{\text{SC}} = \int d^4 x \, \left( \bar{\psi}_s (i \slashed{\del} -\tilde{m}_s v_F^2 +m_s v_F^2 \gamma^0) \psi_s + g \, \psi_s^T \xi_s \, \psi_s \bar{\psi}_s \bar{\xi}_s \bar{\psi}_s^T\right) \, .
\end{equation}
In the introduced action, $\xi_s$, with $\bar{\xi}_s = \gamma^0 \xi_s^\dagger \gamma^0$, is a pairing matrix defined as $\xi_s = i \sigma_y /2 \otimes I_2$ such that we obtain the Cooper pair in the form of $g \langle \psi^T_s \xi_s \, \psi_s \rangle$ in the mean field approximation. The standard BCS action is obtained in the limit  $\sqrt{|\vec{k}^2 - k_F^2|} \ll m_s v_F$ by decomposing the field $\psi_s$ into its large and small components.

Using a Hubbard-Stratonovich transformation, we write the path integral, \( Z[A] = \int D (\bar{\psi}_s, \psi_s) \, e^{i S_{\text{SC}}} \) as
\begin{equation}
Z[A] = \int D (\bar{\psi}_s,\psi_s) D (\bar{\phi},\phi) \, \exp\left( i \int d^4 x \left[ \bar{\psi}_s (i \slashed{\del} -\tilde{m}_s v_F^2 +m_s v_F^2 \gamma^0 ) \psi_s  - \frac{1}{g}|\phi|^2 + \left( \bar{\phi} (\psi_s^T \xi_s \, \psi_s) + \text{h.c.} \right)  \right]  \right) \, .
\end{equation}
If we further introduce the Nambu spinors
\begin{equation}
\bar{\Psi}_s = \begin{pmatrix}
    \bar{\psi}_s & \psi^T_s 
\end{pmatrix}  \, , \quad \Psi_s = \begin{pmatrix}
    \psi_s \\ \bar{\psi}^T_s
\end{pmatrix} \, ,
\end{equation}
the path integral can be expressed as
\begin{equation}
Z[A] = \int D[\bar{\phi},\phi] \exp\left( - i \int d^4 x\, \frac{1}{g} |\phi|^2  \right)  \int D[\bar{\Psi}_s,\Psi_s] \exp\left( i \int d^4 x\,  \bar{\Psi}_s M_s \Psi_s \right) \, ,
\end{equation}
where
\begin{equation}
M_s = \begin{pmatrix}
    ( i \gamma^\mu \del_\mu - \tilde{m}_s v_F^2 + m_s v_F^2 \gamma^0)/2 & \phi \, \bar{\xi}_s    \vspace{0.2cm}
\\
    \bar{\phi} \, \xi_s &  (i \gamma^{\mu \, T} \del_\mu + \tilde{m}_s v_F^2 - m_s v_F^2 \gamma^0)/2
\end{pmatrix} \, ,
\end{equation}
which could be called a relativistic Gor'kov matrix. In the Gor'kov matrix, the diagonal terms emerge as the Dirac Lagrangian can be written as
\begin{equation}
\bar{\psi}_s (i \gamma^\mu \del_\mu -\tilde{m}_s v_F^2 +m_s v_F^2 \gamma^0 ) \psi_s = \frac{1}{2} \bar{\psi}_s (i \gamma^\mu \del_\mu -\tilde{m}_s v_F^2 +m_s v_F^2 \gamma^0) \psi_s - \frac{1}{2} \psi^T_s (-i \gamma^{\mu T} \del_\mu - \tilde{m}_s v_F^2 +m_s v_F^2 \gamma^0 ) \bar{\psi}^T_s \, .
\end{equation}
The fermionic path integral now becomes quadratic in the Nambu basis so that one can perform the integration,
\begin{equation}
  \int D[\bar{\Psi}_s,\Psi_s] \exp\left( i \int d^4 x\,  \bar{\Psi}_s M_s \Psi_s \right) = N \det M_s \, .
\end{equation}
The matrix $M_s$ can be split as \( M_s = S_0^{-1} \left(1 - i S_0 \Phi \right) \) such that $- i S_0^{-1} S_0 = 1$, with 
\begin{equation}
S_0^{-1} =  \frac{1}{2}\begin{pmatrix}
     i \gamma^\mu \del_\mu - \tilde{m}_s v_F^2 + m_s v_F^2 \gamma^0 & 0  \\
    0 & i \gamma^{\mu \, T} \del_\mu + \tilde{m}_s v_F^2 - m_s v_F^2 \gamma^0
\end{pmatrix} \, ,
\end{equation}
and
\begin{equation}
 S_0 (x-y) = 2 \begin{pmatrix}
    G_0 (x-y, m_s)  & 0  \\
    0 &  G_0^T(x-y, -m_s)  
\end{pmatrix} \, , \quad \Phi = \begin{pmatrix}
    0 & \phi \, \bar{\xi}_s  \\
    \bar{\phi} \, \xi_s &  0
\end{pmatrix} \, .
\end{equation}
Here, the Feynman propagator reads
\begin{equation}
G_0 (x-y, m_s) = \int d^4 p \, G_F (p,m_s)e^{-i p (x-y)}, \qquad     G_F (p,m_s) = \frac{i(\slashed{p}+\tilde{m}_s v_F^2 + m_s v_F^2 \gamma^0 )}{(p^0 + m_s v_F^2)^2 - \vec{p}^2 - (\tilde{m}_s v_F^2)^2 +i 0^+}  \, ,
\end{equation}
which follows from
\begin{equation}
(i \slashed{\del}_x - \tilde{m}_s v_F^2 + m_s v_F^2 \gamma^0 )_{\alpha \beta}  \left[ G_0 (x-y,m_s) \right]_{\beta \gamma} =  i \delta_{\alpha \gamma} \delta^{(4)} (x-y) \, .
\end{equation}

After setting the normalization $N = \left( \det S_0^{-1} \right)^{-1}$, the path integral can be expressed as
\begin{equation}
\label{fermionic_part_for_SC}
\int D[\bar{\Psi}_s,\Psi_s] \exp\left( i \int d^4 x\,  \bar{\Psi}_s M_s \Psi_s \right) = \exp\left(- \sum_{n=1} \frac{1}{n} \Tr\left( i S_0 \Phi \right)^n \right) \, ,
\end{equation}
which is given in the leading terms by
\begin{align}
\exp\left(- \sum_{n=1} \frac{1}{n} \Tr\left(i S_0 \Phi \right)^n \right) & = \exp\left( i \int d^4 p \,  \tilde{\phi}(-p) \bar{\tilde{\phi}}(p)  \Pi_1 (p) \right. \\
\notag & \left. + \, i \int d^4 p_1 d^4 p_2 d^4 p_3 \, \tilde{\phi}(-p_1) \bar{\tilde{\phi}}(-p_2) \tilde{\phi}(-p_3) \bar{\tilde{\phi}}(p_1+p_2+p_3) \Pi_2 (p_1,p_2,p_3)
 + \cdots \right) \, . 
\end{align}
Here, $\tilde{\phi}(p)$ is the Fourier transform of the Hubbard-Stratonovich field $\phi(x)$, and 
\begin{align}
\Pi_1 (p) & = - 4 i \int d^4 k \, \Tr \left[ \bar{\xi}_s G_F^T (k-p, -m_s) \xi_s G_F (k, m_s) \right] \, ,  \\
\notag \Pi_2 (p_1,p_2,p_3) & = 8 i \int d^4 k \, \Tr \left[ \bar{\xi}_s G_F^T (k-p_1, -m_s) \xi_s G_F (k-p_1-p_2, m_s) \bar{\xi}_s G_F^T (k-p_1-p_2 -p_3, -m_s) \xi_s G_F (k, m_s)\right] \, ,
\end{align}
are the corresponding Fermion loops.

In the gradient expansion, these loops can be written as
\begin{equation}
\Pi_1(p)  = \Pi_1 (0) + \frac{p^2}{2}  \Pi_1''(0) + \cdots , \qquad
 \Pi_2 (p_1,p_2,p_3) = \Pi_2 (0) + \cdots
\end{equation}
Then, the overall path integral can be expressed as
\begin{equation}
Z[A] = \int D (\bar{\phi},\phi) \, \exp \left( i \int d^4 x \left(  \frac{\Pi_1''(0)}{2} |\del_ \mu \phi|^2 +  \left( - \frac{1}{g} + \Pi_1(0) \right) |\phi|^2  + \Pi_2 (0) |\phi|^4 + \cdots \right) \right) \, .
\end{equation}
The action, up to the quartic order in $|\phi|$, is called the Ginzburg-Landau action, and is written in a concise form as
\begin{equation}
S_{\rm GL} =  \int d^4 x \left( | \del_\mu \phi|^2  - \nu \left(|\phi|^2 - \frac{\delta^2}{2} \right)^2 \right)\, ,
\end{equation}
where we identify
\begin{equation}
\label{delta_id}
    \nu  = -\frac{2 \Pi_2(0)}{\Pi_1''(0)} \, , \quad    \delta  = \sqrt{\frac{1/g - \Pi_1(0)}{\Pi_2 (0)}} \, .
\end{equation}

If we further take into account the presence of a $U(1)$ gauge field $A^\mu$, the Ginzburg-Landau action can be recognized as the Abelian Higgs (AH) model:
\begin{equation}
\label{abelian_higgs}
S_{\text{AH}} =  \int d^4 x \left( |( \del_\mu + i q A_\mu )\phi|^2  - \nu \left(|\phi|^2 - \frac{\delta^2}{2} \right)^2  - \frac{1}{4} F_{\mu \nu} F^{\mu \nu}  \right)\, ,
\end{equation}
with the charge $q = 2 e$. Here, the last term is the usual Maxwell action, with $F_{\mu \nu} = \partial_\mu A_\nu - \partial_\nu A_\mu$.

\subsection{The SC-CI Coupling and The Schrieffer-Wolff Transformation for Induced Pairing}

In the following we consider the pair-pair interaction between the SC and the CI electrons:
\begin{equation}
\label{sc_coupling}
S_{\text{int}} (\psi_s, \psi_q ) = \eta \int d^4 x \, \delta(z) ( \psi_s^T \xi_s \psi_s ) ( \bar{\psi}_q \bar{\xi}_q \bar{\psi}^T_q)  + {\rm h.c.} \, ,
\end{equation}
where $\xi_s$ and $\xi_q = i \sigma_y/2$ represent the pairing matrices for the SC and the CI, respectively, $\eta$ denotes the interaction strength, which we assume $\eta \ll g$, and the Dirac delta, $\delta(z)$, confines the superconducting term to the $x-y$ plane.

To derive the induced pair-pair interaction, we model electron tunneling between the CI system and the SC in the Hamiltonian formalism by writing $H = H_0 + V$, with $H_0 = H_\sM + H_\SC$, where
\begin{equation}
    H_\sM = \sum_{k \sigma} \epsilon_q (k) c^\dagger_{k \sigma} c_{k \sigma}, \qquad H_\SC = \sum_{k \sigma} \epsilon_s (k) d^\dagger_{k \sigma} d_{k \sigma} + \sum_{k} \left( \Delta(k) d^\dagger_{k \uparrow} d^\dagger_{-k\downarrow} + \hc \right) \, ,
\end{equation}
describe the CI and SC systems, respectively. The interaction term, on the other hand,
\begin{align}
    V &= \sum_{k \sigma} \gamma(k) c^\dagger_{k\sigma} d_{k\sigma} + \hc
\end{align}
couples the CI and SC electrons via momentum- and spin-conserving tunneling with the amplitude $\gamma(k)$. 

Next, we apply a Schrieffer-Wolff transformation, 
\begin{align}
    e^{i W} (H_0 + V) e^{-iW} = H_0 + V + i [W, H_0] + i [W, V] + \frac{i^2}{2} [W, [W, H_0]] + \frac{i^2}{2} [W, [W, V]] + O(\gamma^3) \, ,
\end{align}
such that it eliminates the tunneling term at the first order in $\gamma$ by $ [W,H_0] = i V$, and gives rise to the pair-pair interaction via
\begin{equation}
    H_{\rm eff} \equiv  H_0 + \frac{i}{2} [W, V]  + O(\gamma^3) \, .
\end{equation}

Accordingly, after we make the following ansatz,
\begin{equation}
        W = \sum_{k\sigma} A(k,\sigma) c^\dagger_{k\sigma} d_{k\sigma} + \sum_{k\sigma} B(k,\sigma) c^\dagger_{k \sigma} d^\dagger_{-k \sigbar} + \hc \, ,
\end{equation}
where $\sigbar = -\sigma$, we obtain
\begin{equation}\label{eq:AB}
    A_{k,\sigma} =
    \frac{i \gamma(k) (\epsilon_q (k) + \epsilon_s (k) )}{|\Delta(\sigma k)|^2 - \epsilon_q(k)^2 + \epsilon_s(k)^2 } \, , \quad B_{q\sigma}=
    \frac{i \gamma(k) \sigma \Delta(\sigma k)}{|\Delta(\sigma k)|^2 - \epsilon_q(k)^2 + \epsilon_s(k)^2} \, .
\end{equation}
Therefore, the effective Hamiltonian in the leading terms can be written as
\begin{equation}
  H_{\rm eff} \approx  H_0   + \frac{1}{2} \sum_{k \sigma} \frac{|\gamma(k)|^2 \sigma \Delta(\sigma k)}{|\Delta(\sigma k)|^2 - \epsilon_q(k)^2 + \epsilon_s(k)^2} c^\dagger_{k \sigma} c^\dagger_{-k \bar{\sigma}}  + \hc \, ,
\end{equation}
where we assume $\gamma(-k) = \gamma(k)^*$ for simplicity.

Finally, noting that in the mean-field approximation $\Delta(k)= g \braket{d_{-k\sigbar} d_{k\sigma}}$, we deduce that the derived term would correspond to a mean-field approximation of
\begin{equation}
    H_{\rm eff}^{\rm pair} =
    \frac{1}{2} \sum_{k \sigma} \frac{|\gamma(k)|^2 \sigma g}{|\Delta(\sigma k)|^2 - \epsilon_q (k)^2 + \epsilon_s(k)^2} d_{-k\sigbar} d_{k\sigma}c^\dagger_{k \sigma} c^\dagger_{-k \bar{\sigma}} + \hc \, ,
\end{equation}
which leads to the pair-pair interaction introduced in Eq.~\eqref{sc_coupling}. Consequently, we identify \( \eta \sim g \, \gamma_T^2/ \Delta^2 \), with the tunneling amplitude $\gamma_T$ and the superconducting gap $\Delta$.

\subsection{The Total Action}

The total action which describes the SC-CI interface, then, can be expressed as
\begin{align}
S & = \int d^3 x \, \bar{\psi}_q (i \slashed{D} -m_q v_F^2)\psi_q \\ 
\nonumber & + \int d^4 x \left( \bar{\psi}_s (i \slashed{D} -\tilde{m}_s v_F^2 + m_s v_F^2 \gamma^0)\psi_s - \frac{1}{4} F_{\mu \nu} F^{\mu \nu}  + g \, \psi_s^T \xi_s \psi_s \bar{\psi}_s \bar{\xi}_s  \bar{\psi}^T_s  + \eta  \delta(z) \left(  \psi_s^T \xi_s \psi_s \, \bar{\psi}_q \bar{\xi}_q \bar{\psi}^T_q   + {\rm h.c.} \right)\right) \, ,
\end{align}
which can be further written as
\begin{align}
S & = \int d^3 x \, \left(\bar{\psi}_q (i \slashed{D} -m_q v_F^2)\psi_q - \frac{\eta^2}{g L_{\rm CI}} \psi_q^T \xi_q \psi_q \bar{\psi}_q \bar{\xi}_q \bar{\psi}_q^T \right) \\
\nonumber & + \int d^4 x \, \left( \bar{\psi}_s (i \slashed{D} -\tilde{m}_s v_F^2 + m_s v_F^2 \gamma^0 )\psi_s + g \left| \psi_s^T \xi_s \psi_s +  \frac{\eta}{g} \delta(z) \psi_q^T \xi_q \psi_q \right|^2  - \frac{1}{4} F_{\mu \nu} F^{\mu \nu} \right) \, .
\end{align}
Note that after we take square of the relevant term, there emerges $\delta (z)^2$, which can be defined as
\begin{equation}
\delta (z)^2 = \delta (z) \delta (0) \approx \delta (z)/L_{\rm CI} \, ,
\end{equation}
where $L_{\rm CI}$ can be interpreted as the thickness of the CI layer. This follows from the fact that in reality we would define the interaction as
\begin{equation}
S_{\text{int}}^{\text{real}} (\psi_s, \psi_q ) =  \int d^4 x \, \tilde{\eta}(z, L_{\rm CI}) ( \psi_s^T \xi_s \psi_s ) ( \bar{\psi}_q \bar{\xi}_q \bar{\psi}^T_q)  + {\rm h.c.}, \qquad \tilde{\eta}(z, L_{\rm CI}) = \begin{cases} 
      \eta (z) & 0 \leq z \leq L_{\rm CI}\, , \\
      0 & z > L_{\rm CI} \, .
   \end{cases}
\end{equation}
Furthermore, since $\lim_{L_{\rm CI} \to 0^+} \tilde{\eta} (z, L_{\rm CI}) = \eta \delta(z) $ should hold, we can interpret $\delta(0) \approx 1 / L_{\rm CI}$.

\begin{figure}[t]
  \includegraphics[width=0.5\linewidth]{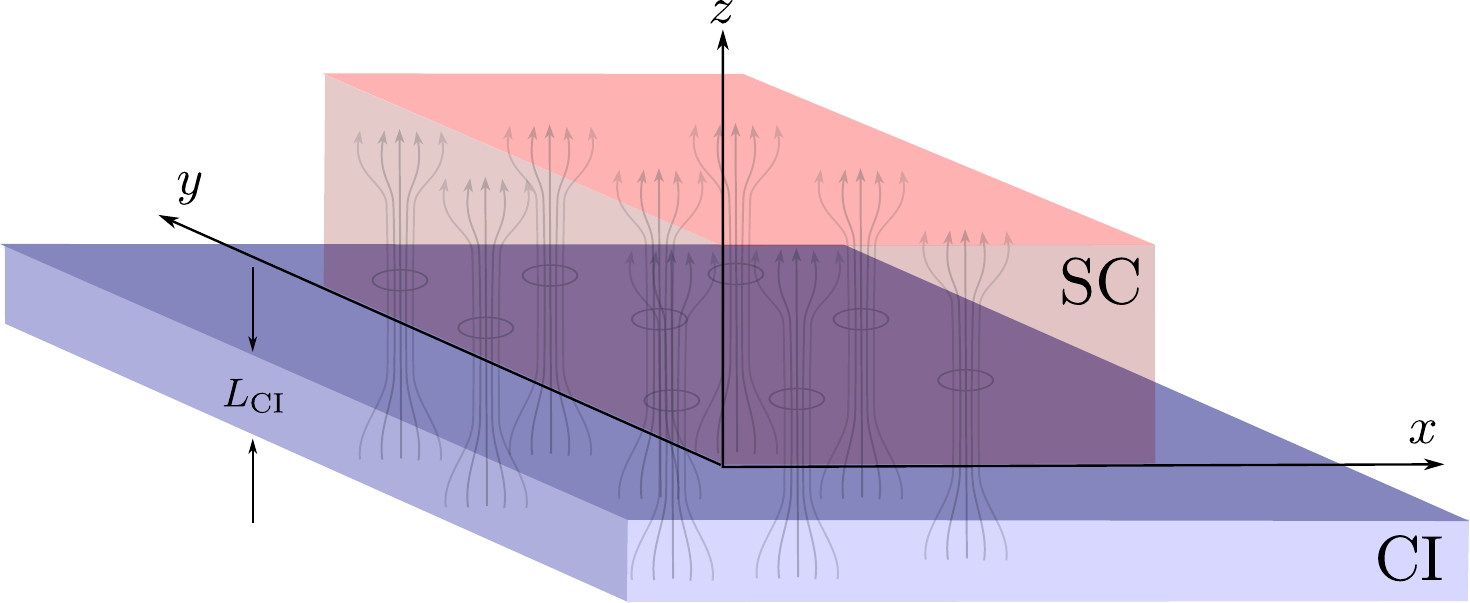}
  \vspace{-0.2cm}
\caption{Geometry of the SC-CI system.}
\vspace{-0.3cm}
\end{figure}

Then, we introduce two auxiliary fields $\phi_1 (t,x,y,z)$ and $\phi_2 (t,x,y)$ such that the path integral can be written as
\begin{align}
& Z[A] = \int D(\bar{\phi}_1,\phi_1) D(\bar{\phi}_2,\phi_2) \exp \left( - i \int d^4 x \, \frac{1}{g} |\phi_1|^2 + i \int d^3 x \, \frac{g L_{\rm CI}}{\eta^2} |\phi_2|^2 - i \int d^4 x \, \frac{1}{4} F_{\mu \nu} F^{\mu \nu} \right) \\
\nonumber & \times \int D(\bar{\psi}_q, \psi_q) D(\bar{\psi}_s , \psi_s) \exp \left( i \int d^3 x \left( \bar{\psi}_q (i \slashed{D}-m_q v_F^2)\psi_q  + \bar{\phi}_2 \psi_q^T \xi_q \psi_q +  \phi_2 \bar{\psi}_q \bar{\xi}_q \bar{\psi}_q^T  \right)   \right) \\
\nonumber \times & \exp \left( i \int d^4 x \left[ \bar{\psi}_s (i \slashed{D} - \tilde{m}_s v_F^2 + m_s v_F^2 \gamma^0) \psi_s  + \bar{\phi}_1 \left(\psi_s^T \xi_s \psi_s + \delta(z) \frac{\eta}{g} \psi_q^T \xi_q \psi_q \right) + \phi_1 \left( \bar{\psi}_s \bar{\xi}_s  \bar{\psi}_s^T + \delta(z) \frac{\eta}{g} \bar{\psi}_q \bar{\xi}_q \bar{\psi}_q^T \right) \right] \right)  \, .
\end{align}
We would like to note here that if we calculate the path integral via the saddle point approximation, we end up with
\begin{equation}
\phi_{1 \, \text{spa}}  = \phi_{1 \, \text{MF}} = g \braket{\psi_s^T \xi_s \psi_s + \delta(z) \frac{\eta}{g} \psi_q^T \xi_q \psi_q } \approx g \braket{ \psi_s^T \xi_s \psi_s}, \qquad \phi_{2 \, \text{spa}}  = \phi_{2 \, \text{MF}} = \frac{\eta^2}{g L_{\rm CI}} \braket{ \psi_q^T \xi_q \psi_q} \, .
\end{equation}

Based on the fermionic part of the path integral, we introduce the following two Nambu spinors
\begin{equation}
\bar{\Psi}_q = \begin{pmatrix}
    \bar{\psi}_q & \psi_q^T  
\end{pmatrix}  \, , \quad \Psi_q = \begin{pmatrix}
    \psi_q \vspace{0.2cm} \\ \bar{\psi}_q^T \vspace{0.1cm} 
\end{pmatrix} \, , \quad \bar{\Psi}_s = \begin{pmatrix} \bar{\psi}_s & \psi_s^T 
\end{pmatrix}  \, , \quad \Psi_s = \begin{pmatrix}
     \psi_s \vspace{0.2cm} \\ \bar{\psi}_s^T
\end{pmatrix} \, ,
\end{equation}
which allows us to define the fermionic part of the path integral as
\begin{equation}
\int D(\bar{\Psi}_q,\Psi_q) e^{i \int d^3 x \, \bar{\Psi}_q M_q \Psi_q} \int D(\bar{\Psi}_s,\Psi_s) e^{ i \int d^4 x \, \bar{\Psi}_s M_s \Psi_s}  =  N \det M_q \det M_s \, ,
\end{equation}
where the matrices can be written as
\begin{subequations}
\begin{align}
M_q & =  \begin{pmatrix}
    ( i \gamma^\mu D_\mu  - m_q v_F^2)/2  &  \bar{\xi}_q (\phi_2 + \eta \phi_1  (0)  /g ) \vspace{0.3cm}\\
    \xi_q (\bar{\phi}_2 + \eta \bar{\phi}_1  (0)/g) &  (  i \gamma^{\mu T} D_\mu^*  + m_q v_F^2)/2   
\end{pmatrix} \quad \text{with} \quad  \phi_1 (0) = \left. \phi_1 \right|_{z=0} \, , \\
\nonumber & \\
M_s & =  \begin{pmatrix}
     (i \gamma^\mu D_\mu  - \tilde{m}_s v_F^2 + m_s v_F^2 \gamma^0)/2  &  \phi_1 \bar{\xi}_s  \vspace{0.2cm}\\
    \bar{\phi}_1 \xi_s & ( i \gamma^{\mu T} D_\mu^* + \tilde{m}_s v_F^2 - m_s v_F^2 \gamma^0)/2  
\end{pmatrix}\, .
\end{align}
\end{subequations}

Therefore, building on the previous sections, where we derived the CS term and the Abelian Higgs model, and further reducing the spatial dimensions to two by assuming translational symmetry along the $z$-direction:
\begin{equation}
\phi_1 (t,x,y,z) =  \phi_1 (t,x,y), \qquad A^\mu (t,x,y,z) = (A_0 (t,x,y), A_x (t,x,y), A_y (t,x,y), 0) \, ,
\end{equation}
we obtain 
\begin{equation}
Z[A] = \int D(\bar{\phi}_1,\phi_1) D(\bar{\phi}_2,\phi_2) e^{i S_{\rm eff}} \, , 
\end{equation}
with the following effective action, which we call the coupled Abelian Higgs-Chern-Simons (HCS) action,
\begin{align}
\label{eff_action_HCS}
S_{\rm eff} = S_{\rm HCS} & = \int d^3 x \, \sum_{j=1,2}\left( |(\del_\mu + i q A_\mu) \phi_j|^2  - \nu_j \left(|\phi_j|^2 - \frac{\delta_j^2}{2}\right)^2 \right) - \beta \int d^3 x \, \left( \bar{\phi}_2 \phi_1 + \phi_2 \bar{\phi}_1 \right) \\
\nonumber & - \frac{1}{4} \int d^3 x\,  F_{\mu \nu}F^{\mu \nu}  + \frac{e^2}{4\pi} \int d^3 x \, \varepsilon^{\mu \nu \rho} A_\mu \del_\nu A_\rho \, ,
\end{align}
with $q = 2 e$ and $\beta = L_{\rm CI}/\eta$. The last term in the first line, which corresponds to the Josephson coupling, arises from the $|\phi_2|^2$ term after the field redefinition $\phi_2 + \eta \phi_1 (0)/g \to \phi_2$ in the path integral.

In the HCS action~\eqref{eff_action_HCS}, $\delta_1$ characterizes the usual superconducting gap of the SC sector, while $\delta_2$ arises from pair-pair interactions and represents the induced superconducting gap in the CI layer. The latter can be given by
\begin{equation}
  \delta_2  = \sqrt{\frac{-g L_{\rm CI}/\eta^2 - \Pi_1^q(0)}{\Pi_2^q (0)}} \, ,
\end{equation}
where 
\begin{equation}
    \Pi_1^q (0) =  - 4 i \int d^3 k \, \Tr\frac{\bar{\xi}_q i(\slashed{k}^T -m_q v_F^2) \xi_q i(\slashed{k} + m_q v_F^2) }{(k^2 - m_q^2 v_F^4 + i 0^+)^2} , \qquad \Pi_2^q (0) = 8 i \int d^3 k \, \Tr \frac{\left(\bar{\xi}_q i (\slashed{k}^T -m_q v_F^2) \xi_q i (\slashed{k} + m_q v_F^2) \right)^2 }{(k^2 - m_q^2 v_F^4 + i 0^+)^4}  \, .
\end{equation}

In the absence of a chemical potential in the CI sector, $\Pi_2^q (0)$ vanishes unless one introduces a momentum cutoff. This reflects the fact that the density of states is zero within the mass gap, and consequently, no spontenous symmetry breaking (SSB), and hence no superconductivity occurs. By introducing a chemical potential $\mu_q \ge m_q v_F^2$, which shifts $k_0 \to k_0 + \mu_q$, a nonzero density of states at the Fermi level emerges. In particular, by setting $\mu_q =  m_q v_F^2$, we obtain
\begin{equation}
    \Pi_1^q(0) = - \frac{m_q v_F^2}{2\pi} , \qquad  \Pi_2^q(0) =  - \frac{1}{8 \pi m_q v_F^2}  \, ,
\end{equation}
and hence the minimum of the Higgs potential yields
\begin{equation}
    \delta_2 = \sqrt{8 \pi m_q v_F^2 \left( \frac{g L_{\rm CI}}{\eta^2} - \frac{m_q v_F^2}{2 \pi} \right)} \, .
\end{equation}

Remark that in the presence of the chemical potential, the polarization tensor~\eqref{polarization_tensor} can be written as
\begin{equation}
    \Pi^{i j} (p) =  \frac{ 1 }{4 \pi} \frac{m_q v_F^2}{|\mu_q|}\varepsilon^{i 0 j} p_0 + \delta_{i j}   \frac{i}{4 \pi} \frac{\mu_q^2-(m_q v_F^2)^2}{\mu_q}\, ,
\end{equation}
where the second term correspond to the longitudinal Drude current. This result emerges as a matter of the fact that upon evaluating the polarization tensor, a Wick rotation crosses a pole when $|\mu_q| > m_q$, therefore one must take into account the additional residue from the crossed pole. Nevertheless, in the particular choice $\mu_q =  m_q v_F^2$, the polarization tensor reduces to the previously obtained result~\eqref{polarization_no_chemical}, and in the absence of the longitudinal current, we still remain in the insulator regime.

\subsection{Spontenous Symmetry Breaking}

We expand the fields $\phi_j$ around their vacuum expectation values as,
\begin{equation}
\phi_j = \frac{1}{\sqrt{2}} (\delta_j + H_j(x)) e^{i \Theta_j (x)} \, .
\end{equation}

Then, the first line of the HCS Lagrangian density can be written as
\begin{equation}
    \mathcal{L} = \sum_j \left[ \frac{1}{2} (\del_\mu H_j)^2 - \nu_j \left( \delta_j H_j + \frac{H_j^2}{2} \right)^2 + \frac{1}{2} (\delta_j + H_j)^2(q A_\mu + \del_\mu \Theta_j)^2 \right] - \beta(\delta_1 + H_1)(\delta_2 + H_2) \cos(\Theta_1 - \Theta_2) \, .
\end{equation}
We further introduce
\begin{equation}
    G = \frac{\delta_1^2 \Theta_1 + \delta_2^2 \Theta_2}{\delta^2}, \qquad P = \frac{\delta_1 \, \delta_2}{\delta} (\Theta_1 - \Theta_2), \quad \text{with } \delta^2 = \delta_1^2 + \delta_2^2 \, .
\end{equation}
Treating $H_j$ as small amplitude fluctuations around $|\phi_j| = \delta_j/\sqrt{2}$ and expanding to quadratic order, we obtain
\begin{equation}
    \mathcal{L} = \sum_j \left[ \frac{1}{2} (\del_\mu H_j)^2 - \nu_j \delta_j^2 H_j^2 \right] + \frac{1}{2} \delta^2 \left(\del_\mu G + q A_\mu \right)^2 + \frac{1}{2} (\del_\mu P)^2 - \beta \delta_1 \delta_2 \cos \left( \frac{\delta P}{\delta_1 \delta_2} \right) + \cdots \, .
\end{equation}
Here, $H_{1,2}(x)$ correspond to two Higgs modes, $G(x)$ is the Goldstone mode, and $P(x)$ is the Leggett mode. Using the gauge transformation $A_\mu \to A_\mu - \partial_\mu G / q$, known as the unitary gauge, the massless Goldstone mode is removed. Expanding $\cos ( \delta P/(\delta_1 \delta_2) )$ around its minimum, the HCS Lagrangian density then can be expressed as
\begin{align}
\mathcal{L}_{\rm HCS} & = \sum_{j=1}^2 \left[ - \frac{1}{2}  H_j \del_\mu \del^\mu H_j - \nu_j \delta_j^2 H_j^2 \right]  -  \frac{1}{4} F_{\mu \nu}F^{\mu \nu} +  \frac{k}{4 \pi} \varepsilon^{\mu \nu \rho} A_\mu \del_\nu A_\rho
+\frac{1}{2} q^2 \delta^2 A_\mu A^\mu \\
& - \frac{1}{2} P \del_\mu \del^\mu P - \frac{1}{2} \frac{\beta \delta^2}{\delta_1 \delta_2} P^2 + \beta  \delta_1 \delta_2 +  \mathcal{L}_{\rm HCS}^{\rm int} \, ,
\end{align}
where $\mathcal{L}_{\rm HCS}^{\rm int}$ describes the interaction among the Higgs modes, the Leggett mode, and massive gauge field. The corresponding masses are $m_{H_j} = \delta_j \sqrt{2 \nu_j} $, $m_{P} = \delta \sqrt{\beta/(\delta_1 \delta_2)}$, and $m_{G_\pm} = \sqrt{q^2 \delta^2 + k^2/16 \pi^2} \pm k/4 \pi $.

\subsection{Vortex Equations}

The equations of motion of the HCS action are then given by
\begin{subequations}
\begin{align}
 \frac{\delta \mathcal{L}_{\rm HCS}}{\delta \bar{\phi}_1} - \del_\mu \frac{\delta \mathcal{L}_{\rm HCS}}{\delta \del_\mu \bar{\phi}_1} & = -  D_\mu D^\mu \phi_1 - \frac{\del V}{\del \bar{\phi}_1} - \beta \phi_2 = 0 \, ,\\
\frac{\delta \mathcal{L}_{\rm HCS}}{\delta \bar{\phi}_2} - \del_\mu \frac{\delta \mathcal{L}_{\rm HCS}}{\delta \del_\mu \bar{\phi}_2} & = - D_\mu D^\mu \phi_2 - \frac{\del V}{\del \bar{\phi}_2} - \beta \phi_1 = 0 \, , \\
 \frac{\delta \mathcal{L}_{\rm HCS}}{\delta A_\nu} - \del_\mu \frac{\delta \mathcal{L}_{\rm HCS}}{\delta \del_\mu A_\nu} & =  \del_\mu F^{\mu \nu} + \frac{k}{4 \pi} \epsilon^{\nu \mu \rho} F_{\mu \rho} - J^\nu = 0 \, ,
\end{align}
\end{subequations}
where
\begin{equation}
J^\nu = i q \left( \phi_1^* (D^\nu \phi_1) -(D^\nu \phi_1)^* \phi_1  +  \phi_2^* (D^\nu \phi_2) -(D^\nu \phi_2)^* \phi_2   \right) \, ,
\end{equation}
is the current associated with the two pairs, and
\begin{equation}
V = \nu_1 \left(|\phi_1|^2 - \frac{\delta_1^2}{2}\right)^2 + \nu_2 \left(|\phi_2|^2 - \frac{\delta_2^2}{2} \right)^2 \, .
\end{equation}

For the vortex ansatz, given by
\begin{equation}
\phi_j = e^{i n \theta} f_j (r) \, , \ A_0 = a_0 (r) \, , \ A_r = 0 \, , \ A_\theta  = a(r)/r  \, ,
\end{equation}
the equations of motion can be written as
\begin{subequations}
\label{vortex_sol}
\begin{align}
&
    f_j''(r)  + \frac{1}{r} f_j'(r) - \frac{1}{r^2} f_j(r) \left[n - q a(r) \right]^2 + q^2 f_j(r) a_0(r)^2 
    - 2 \nu_j [f_j^2(r) - \delta_j^2/2] f_j(r) - \beta |\epsilon^{j i}| f_i = 0 \, ,\\     \label{current_eq}
& 
    a''(r) - \frac{1}{r} a'(r) + \frac{k}{2\pi} r a'_0 (r) 
    + 2 q [n - q a(r)] [f_1^2(r) + f_2^2(r)] = 0 \, , \\
    \label{charge_eq}
& 
    a_0''(r) + \frac{1}{r} a_0'(r) + \frac{k}{2\pi r} a'(r) 
    - 2 q^2 a_0 (r) [f_1^2(r) + f_2^2(r)] = 0 \, .
\end{align}
\end{subequations}
From the above equations, we derive the following important results. First, Eq.~\eqref{charge_eq} can be expressed as
\begin{equation}
\vec{\nabla} \cdot \vec{E} + \frac{k}{2\pi} B = j^0  \, ,
\end{equation}
where $j^0 = - 2 q^2 a_0 (r) [f_1^2(r) + f_2^2(r)]$, and it defines the charge of the paired vortex
\begin{equation}
Q  = \int d^2 x \, j_0 = \int d^2 x [\vec{\nabla} \cdot \vec{E} + k B /(2 \pi) ] = \frac{k}{2\pi} \int d^2 x B = \frac{k}{2\pi}  \Phi_n  =  -n \, k /q  = - n \, e/2 \, .
\end{equation}
Next, the current density, governed by Eq.~\eqref{current_eq}, can be written as
\begin{equation}
 j_\theta = - \frac{k}{2 \pi} E_r - \frac{\del}{\del r} B \, .
\end{equation}
Here, the first term is the edge curren, representing the Hall conductivity. The second term, $(\vec{\nabla} \times \vec{B})_\theta$, corresponds to the combined London equation, which defines the supercurrent.

Moreover, the angular momentum carried by the vortices can be calculated via
\begin{equation}
    M_z = \int d^2 x \, \epsilon^{i j} x_i T_{0 j}, \qquad T_{0 j} =2 q n \frac{a_0}{r} (f_1^2 +f_2^2)  \delta_{j \theta}\, .
\end{equation}
Then, using the equation of motion~\eqref{charge_eq}, we obtain
\begin{equation}
    M_z = - \frac{n}{q} Q = \left(\frac{n}{2} \right)^2 \, .
\end{equation}

\subsection{Vortex-Vortex Interaction}

To illustrate the qualitative behavior of the vortex-vortex interaction, we work in the regime where the CI layer is sufficiently thin that the Josephson coupling, characterized by $\beta = L_{\rm CI}/\eta$, can be neglected. Then, the energy functional of the paired vortex is given by
\begin{equation}
\label{gibbs_energy}
 \mathcal{E}  = \int d^2 x \, \left( \vec{E}^2 + B^2 \right) /2  + \int d^2 x \sum_{j=1}^2 \left( q^2 A_0^2 |\phi_j|^2 + |\vec{D} \phi_j|^2 + \nu_j \left(|\phi_j|^2 - \frac{\delta_j^2}{2}\right)^2  \right)  \, .
\end{equation}
Here, we emphasize that the presence of the CS term in the HCS action is reflected in the electric field $\vec{E}$ and its associated scalar potential $A_0$, representing a critical distinction from the standard Ginzburg-Landau energy functional, where $\vec{E} = -\nabla A_0 = 0$.

The energy functional can be written in terms of the introduced ansatz as
\begin{equation}
\mathcal{E}_n (k) = 2 \pi \int_0^\infty dr \, \left( (f_1^2 + f_2^2 ) \left( q^2 r \, a_0^2 + q (n-q \, a) \frac{a}{r} \right) + r \sum_{j=1}^2 \nu_j \left(\frac{\delta_j^4}{4} - f_j^4 \right) \right) \, ,
\end{equation}
which was simplified using the equations of motion.

Now, assuming $k/(2\pi) < 1$, we introduce the following expansions
\begin{equation}
f_j = \sum_{m = 0} g_{j, \, m}(r)\left(\frac{k}{2\pi} \right)^m \, , \ a(r) = \sum_{m=0} \bar{c}_m(r) \left(\frac{k}{2\pi} \right)^m\, , \ a_0(r) = \sum_{m=0} \bar{b}_m(r) \left(\frac{k}{2\pi} \right)^m   \, ,
\end{equation}
where the $m=0$ components correspond to the vortex solution in the absence of the CS term. Furthermore, due to the symmetry of the equations of motion under $k \to -k$, which implies $a_0 \to  - a_0$, we conclude that $g_{j, \,  2m+1} = \bar{c}_{2m + 1} = \bar{b}_{2m} = 0$.

The corresponding coupled equations of motion at the leading order can be solved with the following coefficients:
\begin{equation}
g_{j, \, 2} = 0 \, , \ \bar{c}_2 = - \frac{r^2}{8 q} (n- q \bar{c}_0) \, , \ \bar{b}_{1} = (n- q \bar{c}_0) \frac{1}{2 q} \, .  
\end{equation}
Then, the energy functional can be expressed as
\begin{equation}
\mathcal{E}_n (k) = \mathcal{E}_n(0) + \frac{n^2}{8 q^2} \left(\frac{k}{2\pi} \right)^2 + \cdots
\end{equation}
where $\mathcal{E}_n(0)$ refers to the energy of the chargeless vortex. Then, we can define the asymptotic interaction energy between two vortices as
\begin{equation}
    \Delta \mathcal{E} (k) = \mathcal{E}_n (k) - n \mathcal{E}_1 (k) = \Delta \mathcal{E} (0) + \frac{(n^2 - n)}{8 q^2}  \left(\frac{k}{2\pi} \right)^2 + O \left(\frac{k}{2\pi} \right)^4  \, ,
\end{equation}
where the second term represents an additional repulsion between the charged vortices.

\end{document}